\documentclass[11pt,a4paper]{article}
\usepackage[T1]{fontenc}
\usepackage[utf8]{inputenc}
\usepackage{lmodern}
\usepackage[margin=1in]{geometry}
\usepackage{amsmath,amssymb,mathtools,graphicx}
\usepackage{microtype}
\usepackage{placeins}
\usepackage[colorlinks=true,linkcolor=blue,citecolor=blue,urlcolor=blue]{hyperref}
\graphicspath{{figures/}{./}}
\newcommand{\manuscriptfigure}[2][\linewidth]{\includegraphics[width=#1]{#2}}
\newcommand{\lGB}{\lambda_{\mathrm{GB}}}
\newcommand{\F}{\mathcal{F}}
\newcommand{\Ncal}{\mathcal{N}}
\newcommand{\BH}{\mathrm{BH}}
\newcommand{\MN}{\mathrm{MN}}
\title{A Holographic Weyl Semimetal in an Effective Gauss--Bonnet Geometry}
\author{F. Mousavi\thanks{E-mail: musavi.fatemesadat1994@shahrood.ac.ir}
	\quad and \quad K. Bitaghsir Fadafan\thanks{E-mail: bitaghsir@shahroodut.ac.ir}\\[0.4em]
	\small Faculty of Physics, Shahrood University of Technology,\\
	\small P.O. Box 3619995161, Shahrood, Iran}
\date{}
\begin{document}
	\maketitle
	
\begin{abstract}
We study how an effective Gauss--Bonnet deformation of the thermal
geometry changes the competition between the conducting Weyl-semimetal
and insulating flavour phases of the probe D3/D7 model. The phases are
represented by horizon-reaching and smoothly capped D7 embeddings,
respectively, whose free energies exhibit a first-order transition.
At the temperatures examined, positive Gauss--Bonnet coupling shifts
coexistence toward larger flavour mass and expands the thermodynamically
preferred Weyl-semimetal region, while negative coupling shifts the
boundary in the opposite direction. The normalized zero-temperature
transition remains unchanged. An on-shell analysis shows that the
leading finite-temperature shift is largely explained by an effective
rescaling of the thermal scale, opposed by an intrinsic fixed-horizon
geometric contribution. The enlargement of the conducting phase does
not imply stronger electrical response: the longitudinal and anomalous
Hall conductivity jumps at coexistence can decrease, and the coupling
response of the longitudinal jump reverses with temperature. The
mass-dependent maximum of the longitudinal DC conductivity can also
cease to occur on the thermodynamically preferred branch. Thus the
geometric deformation changes both which phase is favored and how the
Weyl-semimetal phase conducts.
\end{abstract}
\newpage
\tableofcontents

\newpage
\section{Introduction}
\label{sec:intro}

In a three-dimensional Weyl semimetal, two bands touch at isolated
momenta called Weyl nodes. Each node carries a chirality that protects
the crossing against small perturbations. Nodes of opposite chirality
can move together and annihilate, opening a gap. In a system with
broken time-reversal symmetry, their separation in momentum space
can also give rise to an anomalous Hall current in the absence of an
external magnetic field~\cite{ArmitageReview}. This raises a question
when interactions are strong: does a perturbation that extends the
conducting phase also strengthen its electrical response?

Interactions can affect the transition and the current through different
mechanisms. Short-range interactions can drive a first-order transition
to a band insulator or a continuous transition into a symmetry-broken
phase~\cite{RoyInteractingWeyl}. Other field-theory studies find
interaction corrections to anomalous electromagnetic
response~\cite{ParhizkarRylandsGalitski2024} and changes in both the
Weyl nodes and the Hall response in an interacting Weyl--Hubbard
model~\cite{NandyLaneZhu2024}.

To identify the two scales that compete in our model, consider first
a free Dirac fermion of mass $m$ in a spatial axial source $A_z^5=b/2$~\cite{Grushin2012}.
Its two positive-energy branches obey
\begin{equation}
	\varepsilon_\pm^2(\boldsymbol{k})
	=
	k_x^2+k_y^2+
	\left(
	\sqrt{k_z^2+m^2}\pm\frac b2
	\right)^2.
	\label{eq:freeDirac}
\end{equation}
The lower branch has two nodes at $k_z=\pm\sqrt{b^2/4-m^2}$ for
$|m/b|<1/2$. Increasing the mass brings the nodes together until a gap
opens. The ratio $m/b$ therefore gives a natural parameter for tracking
the transition even when a quasiparticle dispersion is unavailable.

In holography one can locate the transition at strong coupling by
comparing free energies and compute the current in the selected phase.
Bottom-up constructions realize Weyl-semimetal transitions
and anomalous transport using axial gauge fields and
Chern--Simons interactions
\cite{HolographicWSM,TopologicalTransition,TopologicalReview}.
Some holographic models also exhibit a first-order transition into a
Chern-insulating phase~\cite{LiuZhao2018}, distinct from the trivial
flavour insulator studied here. The axial Hall effect and its anomaly
relation have been investigated in related holographic
constructions~\cite{CopettiAxialHall2017}; this axial-current
response should not be confused with our vector-current Hall conductivity.
Our starting point is the probe-flavour construction of
Ref.~\cite{Fadafan}, based on D7-branes in the D3-brane
geometry. Its boundary theory is strongly coupled
$\mathcal N=4$ super-Yang--Mills theory with
$N_f\ll N_c$ fundamental
$\mathcal N=2$ hypermultiplets~\cite{Maldacena,KarchKatz}.
A spatially varying mass phase, supplies the
axial source. The D7-branes either reach the horizon
or close smoothly above it, giving black-hole (BH)
and Minkowski (MN) embeddings, respectively. These topologically
distinct probe-brane embeddings and the transitions between them
have a well-established history~\cite{MateosPhase2006}.
A first-order transition separates the conducting
semimetal and insulating flavour phases.

The Einstein D3/D7 model has several distinctive transport
properties. At zero temperature, the Hall conductivity
of the semimetal phase is independent of the flavour mass
and takes its anomaly-normalized value. At finite
temperature, both the longitudinal and Hall conductivities
depend on the embedding. In particular, the longitudinal
DC conductivity develops a maximum as a function of mass
at sufficiently low temperature, while becoming monotonic
at higher temperature~\cite{Fadafan}.
A subsequent calculation of finite-frequency conductivities
and quasinormal modes identified weakly damped current
excitations associated with optical structures near the
transition~\cite{Furukawa2025}. Optical conductivity has also
been studied in bottom-up holographic Weyl semimetals~\cite{GrignaniAC2017}.
Those frequency-dependent structures should be distinguished
from a maximum of the DC conductivity as the mass is varied.

Deformations of the same flavour-brane construction provide
further tests of these properties.
The D-instanton extension changes the dilaton and probe
dynamics, shifts the phase boundary and suppresses the
low-temperature longitudinal DC
maximum~\cite{EomSeo2026}.
We show the same suppression of this maximum by considering
higher-curvature corrections. Its interpretation requires
both the conductivity formula and the thermodynamic
selection of the embedding: an extremum on a stationary
branch need not be reached by the preferred phase.

Here we examine a deformation of the thermal geometry
through a five-dimensional Gauss--Bonnet (GB) interaction.
GB gravity supplies an explicit planar black-brane
background and second-order metric equations, making it
a useful effective model for investigating geometric
corrections. Its thermal response and consistency
conditions have been widely studied
\cite{CaiGB2002,Brigante,BriganteCausality2008,BuchelMyers,CamanhoEdelstein}.
Applications to heavy-quark observables and flavour
conductivity provide additional motivation for studying
the probe sector
\cite{FadafanDragGB,FadafanChargeGB,
	AliAkbariFadafanConductivity,FadafanStrangeGB}.
Gauss--Bonnet--Maxwell models have also been used to study
DC conductivity and charge diffusion~\cite{GeGBMaxwell2012};
their charged bulk gauge sector differs from the neutral D7 probe
flavour sector considered here. This deformation is an effective prescription, not the specific
finite-'t Hooft-coupling correction to type IIB theory, whose leading
known term involves $R^4$ interactions~\cite{GubserKlebanovTseytlin1998}.
A separate Einstein--Gauss--Bonnet holographic construction
addresses torsional dislocations~\cite{JuricicMiskovicCarrasco2025},
not the flavour-brane transition considered here.

We vary the thermal geometry while retaining the D7 action and flavour
dictionary of the Einstein model. The resulting construction is a defined
effective probe prescription, not a ten-dimensional GB uplift. In
particular, the internal $S^5$, RR coupling and relative DBI/WZ
normalization are held fixed; changing these choices could alter the
response. We therefore focus on what this specific geometric deformation
does beyond a shift of the effective thermal scale.

Our central result is the phase diagram in
Fig.~\ref{fig:phase_diagram}. At the sampled finite temperatures, positive
GB coupling extends the preferred semimetal region toward larger mass.
An on-shell variation predicts the leading shift and separates the
thermal-rescaling contribution from an opposing intrinsic fixed-horizon
contribution. At $T/b=0.10$ both conductivity jumps decrease with increasing
coupling, whereas at $T/b=0.07$ the longitudinal jump increases; a finer
scan would be needed to determine the temperature at which the response
changes sign. Matched-horizon tests show that much of the response at
$T/b=0.10$ follows from thermal rescaling, with smaller residual effects.
We also distinguish the kinematic geometric criterion for an interior
longitudinal DC maximum from its realization on thermodynamically
preferred embeddings.

Section~\ref{sec:model} specifies the effective model, temperature and mass
conventions. Section~\ref{sec:embeddings} compares the competing embeddings
and separates the thermal and intrinsic geometric contributions to the
coexistence displacement. Section~\ref{sec:transport} treats the DC
conductivities. We conclude in Section~\ref{sec:conclusions}, with numerical
checks in Appendices~\ref{app:conventions} and~\ref{app:analyticchecks}.

\section{Holographic model with Gauss--Bonnet corrections}\label{sec:model}
The Einstein model starts with $N_c$ coincident D3-branes extending along
the four field-theory directions $(t,x,y,z)$ and $N_f$ D7-branes
introduced as probes, with $N_f\ll N_c$~\cite{KarchKatz,Fadafan}.
In the near-horizon limit, the D3-branes generate the
$\mathrm{AdS}_5\times S^5$ geometry. Strings connecting D3- and
D7-branes supply the fundamental flavour fields, while the D7 embedding
encodes their mass and response. The D7-branes additionally wrap four transverse directions and are
embedded in the remaining $(X^8,X^9)$ plane.

These four directions $(X^4,\ldots,X^7)$ have a radial coordinate
$r_4$ and the angles of a three-sphere. The remaining plane
$(X^8,X^9)$ is described by a radius $r_2$ and an azimuthal angle
$\phi$. To define the polar angle on the five-sphere, write
\begin{equation}
 r_6^2=r_4^2+r_2^2,\qquad
 r_4=r_6\cos\theta,\qquad
 r_2=r_6\sin\theta,\qquad
 X^8+iX^9=r_2e^{i\phi},
 \label{eq:polarcoordinates}
\end{equation}
where $0\leq\theta\leq\pi/2$ and $\phi$ is periodic with period
$2\pi$. The corresponding unit-sphere metric is
\begin{equation}
 d\Omega_5^2=d\theta^2+\cos^2\theta\,d\Omega_3^2
             +\sin^2\theta\,d\phi^2.
 \label{eq:internalgeometry}
\end{equation}
Thus $\theta$ determines the radius of the wrapped $S^3$, whereas
$\phi$ specifies the direction of the brane in the transverse
two-plane~\cite{Fadafan,Furukawa2025}.

Our bulk radial coordinate is $u$, which increases from the AdS
boundary into the interior. In curvature-radius units, the
zero-temperature Einstein metric takes the form
\begin{equation}
 ds_{10}^2=\frac{1}{u^2}
 \left(-dt^2+dx^2+dy^2+dz^2+du^2\right)+d\Omega_5^2.
 \label{eq:vacuumcoordinates}
\end{equation}
Restoring the common Einstein curvature radius gives
$u=L_{\mathrm{eff}}^2/r_6$; Eq.~\eqref{eq:vacuumcoordinates}
uses $L_{\mathrm{eff}}=1$, so $u=1/r_6$ there.
Hence $u=0$ is the UV boundary,
while larger $u$ probes the infrared. At finite temperature the
coordinate $u$ is defined by the metric in
Eq.~\eqref{eq:metric}, with a horizon at $u=u_h$; it should not
be identified with the inverse of an isotropic finite-temperature
Cartesian radius without the appropriate coordinate transformation.

The D7 wraps the $S^3$ and extends along $(t,x,y,z,u)$, so a convenient
worldvolume chart and the equilibrium embedding are
\begin{equation}
 \xi^a=(t,x,y,z,u,\Omega_3),\qquad
 \theta=\theta(u),\qquad \phi=bz.
 \label{eq:embeddingcoordinates}
\end{equation}
Here $\Omega_3$ denotes three independent angles. The fields
$\theta$ and $\phi$ are embedding functions, not additional independent
worldvolume coordinates. The UV coefficient $M$ in $\theta\sim Mu$ specifies the mass source;
see Eq.~\eqref{eq:UV}. The wrapped $S^3$ collapses as
$\theta\to\pi/2$, allowing smooth capped solutions; a horizon-reaching
brane instead gives a BH embedding. We take $b>0$ and $M\geq0$.

The winding angle supplies the axial source independently of the
five-dimensional thermal metric. We therefore keep the internal
geometry fixed and deform the black brane by a GB interaction.
This specifies an effective probe model. No corrected ten-dimensional
type-IIB solution is assumed: the internal $S^5$ radius, the RR
five-form contribution in the probe action, and the flavour mass and
current dictionaries are held fixed by prescription. A genuine
finite-'t Hooft-coupling completion would generally change additional
bulk and worldvolume terms~\cite{GubserKlebanovTseytlin1998}.
To compare thermal states across couplings, we first fix
the normalization of boundary time.

Let $L_0$ and $L_{\mathrm{eff}}$ denote the bare curvature radius and the
asymptotic AdS radius, respectively. The gravitational action is
\begin{equation}
	S_{\mathrm{grav}}
	=\frac{1}{16\pi G_5}\int d^5x\sqrt{-g}
	\left[
	R+\frac{12}{L_0^2}
	+\frac{\lGB L_0^2}{2}
	\left(R_{abcd}R^{abcd}-4R_{ab}R^{ab}+R^2\right)
	\right].
	\label{eq:gravity}
\end{equation}
We use the branch continuously connected to Einstein gravity and define
\begin{equation}
	n=\frac{1+\sqrt{1-4\lGB}}{2},
	\qquad L_{\mathrm{eff}}^2=nL_0^2.
\end{equation}
In units where $L_{\mathrm{eff}}=1$, the black-brane metric is
\begin{equation}
	ds_5^2=\frac{1}{u^2}
	\left[
	-f(u)\,dt^2+d\boldsymbol{x}^2+\frac{du^2}{f(u)}
	\right],
	\qquad
	f(u)=\frac{n}{2\lGB}
	\left[
	1-\sqrt{1-4\lGB
		\left(1-\frac{u^4}{u_h^4}\right)}
	\right].
	\label{eq:metric}
\end{equation}
The Einstein limit is $f(u)=1-u^4/u_h^4$. Since $f(0)=1$,
the boundary speed of light is unity, allowing masses, temperatures
and response functions to be compared in common boundary units.

Near the horizon,
$f(u)=4n(u_h-u)/u_h+\mathcal{O}((u_h-u)^2)$.
Regularity of the Euclidean geometry therefore gives
\begin{equation}
	T=\frac{|f'(u_h)|}{4\pi}=\frac{n}{\pi u_h},
	\qquad
	u_h=\frac{n}{\pi T}.
	\label{eq:temperature}
\end{equation}
We impose this relation separately at each coupling when comparing
states at fixed physical temperature. In particular, positive $\lGB$
decreases $n$ and hence moves the horizon to smaller $u_h$ at fixed $T$.
Comparisons at fixed $u_h$ instead involve different physical
temperatures. The equivalent bare-radius convention is discussed in
Appendix~\ref{app:conventions}.

This normalization also helps interpret the GB dependence: an Einstein
background at $T_{\mathrm{eff}}=T/n$ has the same horizon position as
the GB background at temperature $T$. Consequently, part of the
coupling dependence can reflect a change of the thermal scale.
The blackening functions nevertheless remain different. At fixed
$u_h$ and $v=(u/u_h)^4$, their difference begins at
$f_{\rm GB}-f_{\rm E}=-\lGB v(1-v)+\mathcal O(\lGB^2)$.
Thus matching the horizon position does not make the two embedding
problems exactly equivalent.

For reference, the five-dimensional Gauss--Bonnet causality and
energy-flux bounds are
$-7/36\leq\lGB\leq9/100$~\cite{BuchelMyers,CamanhoEdelstein}.
All sampled couplings lie within this interval, with $\lGB=0.09$
at its upper endpoint. These bounds do not by themselves establish
a UV completion of the combined gravity--probe construction.
	
	With the thermal metric fixed, the D7 embedding and its current response
follow from the DBI and Wess--Zumino actions in the induced geometry,
\begin{align}
		S_{\mathrm{DBI}}&=-N_fT_7\int d^8\xi\sqrt{-\det(P[g]+2\pi\alpha'F)},\\
		S_{\mathrm{WZ}}&=\frac{N_fT_7(2\pi\alpha')^2}{2}
		\int P[C_4]\wedge F\wedge F.
		\label{eq:WZ}
	\end{align}
The equilibrium embedding is given in Eq.~\eqref{eq:embeddingcoordinates},
with vanishing worldvolume field strength $F=0$.
The profile $\phi=bz$ gives the complex flavour-mass source a
spatially varying phase. In the convention of Ref.~\cite{Fadafan}, its
axial source is $A_z^5=b/2$, selecting the $z$ direction and distinguishing
it from the transverse plane. The equilibrium gauge field vanishes, so
the WZ term contributes no background free energy. In the orientation
and RR-potential convention of Ref.~\cite{Fadafan}, the part of the
four-form pullback that enters the Hall response is
\begin{equation}
 P[C_4]_{\mathrm{int}}=-b\cos^4\theta\,
 dz\wedge\omega(S^3),
 \label{eq:fourformpullback}
\end{equation}
where $\omega(S^3)$ is the unit-volume form of the wrapped sphere.
We retain this internal potential when the thermal metric is deformed.
Its contribution to the current appears when a transverse electric field
is applied. Thus the free-energy crossing and Hall response involve
different terms of the probe action, evaluated on the same equilibrium
embedding.
	
After integrating over the wrapped $S^3$ and removing the common
normalization, the Euclidean radial density is
	\begin{equation}
		\mathcal I(u)=\frac{\cos^3\theta}{u^5}
		\sqrt{\bigl(1+b^2u^2\sin^2\theta\bigr)
			\bigl(1+u^2f\theta'^2\bigr)}.
		\label{eq:density}
	\end{equation}
The factor $\cos^3\theta$ measures the size of the wrapped three-sphere.
The two square roots encode the axial-phase gradient along $z$ and the
radial bending of the brane, respectively. Their competition permits
horizon-reaching and smoothly capped configurations. Changing $f$ alters
the cost of radial bending and the horizon position while keeping the
axial source fixed.

We write the flavour free-energy density as $\F=F/\Ncal$; all free
energies and responses below are densities per unit boundary volume.
	The retained Einstein probe dictionary sets
	$\Ncal=\lambda_{\mathrm{tH}}N_fN_c/(16\pi^4)$ and
	$K=\Ncal(2\pi\alpha')^2=N_fN_c/(4\pi^2)$ in the chosen units.
We hold $\lambda_{\mathrm{tH}}$ and the relative DBI and WZ
normalizations fixed as $\lGB$ varies. A common positive multiplier of
the entire probe action would affect absolute current-response
normalizations but not the embedding equations or free-energy crossing.
More generally, multiplying the entire probe action by
an arbitrary positive coupling-dependent factor $A(\lGB)$ leaves the
stationary embeddings and the equal-free-energy condition unchanged:
$\Delta\F\mapsto A(\lGB)\Delta\F$. Such a rescaling changes the absolute
mass response and current normalization, but not the coexistence mass.
Common finite boundary counterterms also cancel between the two branches
at identical sources. These invariances do not cover changes of the
internal geometry or of the relative DBI and WZ couplings, which would
define different effective probe models.

To compare the phases at a common flavour mass, we need the relation
between each regular embedding and its boundary source. Varying the
density in Eq.~\eqref{eq:density} gives the equation for $\theta(u)$.
Its two metric factors are
\begin{equation}
	D=1+b^2u^2\sin^2\theta,
	\qquad
	H=1+u^2f(u)\theta'^2.
\label{eq:embeddingaux}
\end{equation}
Variation of Eq.~\eqref{eq:density} gives
\begin{equation}
	\frac{d}{du}
	\left(
	\frac{f\cos^3\theta\sqrt D}{u^3\sqrt H}\,\theta'
	\right)
	=
	\frac{\cos^3\theta\sqrt{DH}}{u^5}
	\left(
	-3\tan\theta
	+\frac{b^2u^2\sin\theta\cos\theta}{D}
	\right).
	\label{eq:eom}
\end{equation}
Here primes denote derivatives with respect to $u$. The shrinking
three-sphere and the axial winding contribute the two terms on the
right-hand side. The GB deformation changes the radial bending cost
through $f(u)$ and changes the horizon position through
Eq.~\eqref{eq:temperature}.

The relation between a regular embedding and its field-theory sources
is determined near the boundary. Since
\begin{equation}
	f(u)=1-\frac{n}{\sqrt{1-4\lGB}}
	\left(\frac{u}{u_h}\right)^4+\mathcal O(u^8),
	\label{eq:fUV}
\end{equation}
the leading embedding equation is
$u^2\theta''-3u\theta'+3\theta=0$, with solutions proportional to
$u$ and $u^3$. Including the axial contribution gives
\begin{equation}
	\theta(u)
	=
	Mu+
	\left[
	c_3+\frac{b^2M}{2}\log(u\Lambda_{\rm UV})
	\right]u^3+\cdots.
	\label{eq:UV}
\end{equation}
Substitution into the nonlinear equation at order $u^3$ gives
$2a_{\log}-b^2M=0$, where $a_{\log}$ multiplies
$u^3\log(u\Lambda_{\rm UV})$. The cubic terms present at $b=0$ cancel at
this order. Since $f-1=\mathcal O(u^4)$, its first contribution
to the equation multiplied by $u^2$ is of order $u^5$; it cannot
modify the displayed powers or logarithmic coefficient. The GB coupling
nevertheless changes the relation between $M$ and $c_3$ through the
globally regular solution.

We retain the physical mass notation and flavour dictionary of
Ref.~\cite{Fadafan},
\begin{equation}
	\frac{m}{b\sqrt{\lambda_{\mathrm{tH}}}}
	=
	\frac{M}{2\pi b}.
	\label{eq:mass}
\end{equation}
Thus $M$ is the UV embedding coefficient, while $m$ is the physical
flavour mass. We use $c_3$ for the subleading coefficient of
$\theta(u)$ to distinguish it from the coefficient $C$ in the
Cartesian embedding expansion of Ref.~\cite{Fadafan}. Neither
coefficient by itself is the renormalized mass response.

\section{Gauss--Bonnet phase structure}
\label{sec:embeddings}

The bulk equation and UV dictionary are specified in
Section~\ref{sec:model}. We now construct regular
black-hole (BH) and Minkowski (MN) embeddings, compare their free
energies at identical sources, and determine how the resulting
first-order boundary responds to the GB deformation. These steps
must be distinguished: a stationary embedding is not necessarily
the thermodynamically preferred phase.

\label{sec:embeddingfamilies}

The regular embeddings differ in their infrared boundary
conditions. A BH embedding reaches $u_h$ with
$0\leq\theta_h<\pi/2$, so the wrapped three-sphere has nonzero radius
at the horizon. Regularity fixes
\begin{equation}
	\theta'(u_h)
	=
	\frac{
		-3\tan\theta_h+
		\dfrac{b^2u_h^2\sin\theta_h\cos\theta_h}
		{1+b^2u_h^2\sin^2\theta_h}
	}{u_h^2 f'(u_h)},
	\label{eq:horizon}
\end{equation}
where
\begin{equation}
	f'(u)
	=
	-\frac{4nu^3}
	{u_h^4\sqrt{1-4\lGB(1-u^4/u_h^4)}}.
	\label{eq:fprime}
\end{equation}
The induced horizon permits absorption of flavour-current
fluctuations and supplies the dissipative channel relevant to the
DC response.

An MN embedding instead reaches $\theta=\pi/2$ at $u_0<u_h$.
The wrapped three-sphere shrinks smoothly, and the brane closes before
reaching the horizon. A regular local coordinate is
$\rho=\cos\theta$, in which
\begin{equation}
	u(\rho)=u_0+a_2\rho^2+\mathcal O(\rho^4),
	\qquad
	a_2=
	\frac{u_0^2f(u_0)}{8}
	\left[
	-\frac4{u_0}
	+\frac{f'(u_0)}{2f(u_0)}
	+\frac{b^2u_0}{1+b^2u_0^2}
	\right].
	\label{eq:cap}
\end{equation}
The divergence of $\theta'(u)$ at the cap is therefore a coordinate
effect. In the Einstein construction, capped embeddings describe a
discrete, gapped flavour spectrum~\cite{KruczenskiMesons2003,Fadafan,Furukawa2025};
the retained probe action also gives their vanishing DC response
at zero density.

\begin{figure}
	\centering
	\includegraphics[width=0.8\textwidth]{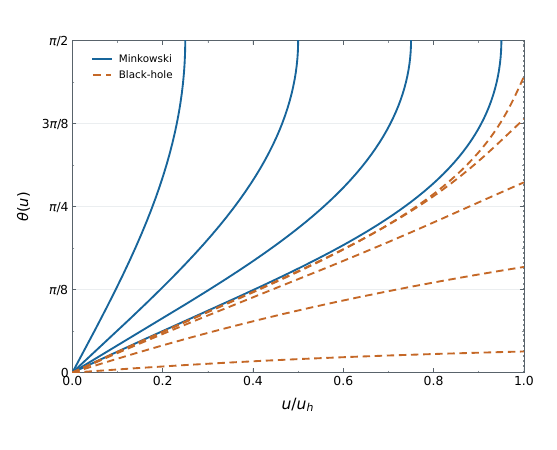}
	\caption{Representative embedding families at $\lGB=0.05$,
		$bu_h=1$, and $T/b\simeq0.30151$. Solid blue curves are
		Minkowski embeddings with $u_0/u_h=0.25,0.50,0.75,0.95$;
		dashed orange curves are black-hole embeddings with
		$\theta_h=0.10,0.50,0.90,1.20,1.40$.
		The MN embeddings close where the wrapped three-sphere
		collapses at $\theta=\pi/2$, whereas the BH embeddings reach
		$u/u_h=1$. The curves have different UV masses and illustrate
		the two regular families before thermodynamic selection.}
	\label{fig:profiles}
	\vspace{9pt}
\end{figure}
\begin{figure}
\manuscriptfigure[0.91\textwidth]{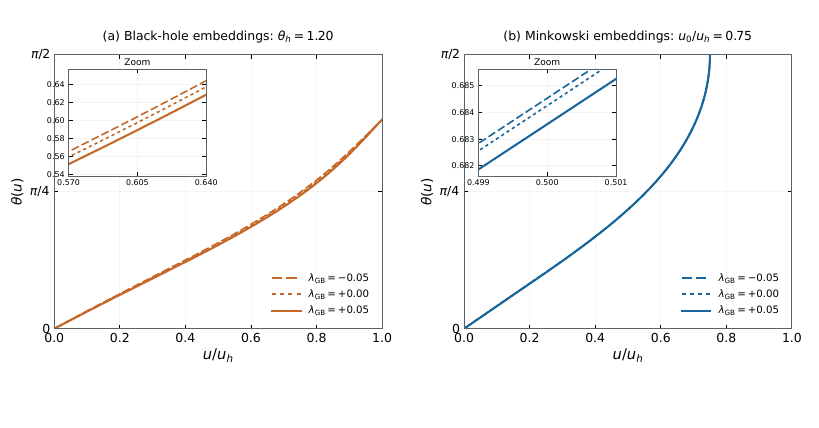}
\caption{GB dependence of representative embeddings at common
		physical temperature $T/b\simeq0.30151$. Panel (a) fixes
		$\theta_h=1.20$ for BH embeddings; panel (b) fixes
		$u_0/u_h=0.75$ for MN embeddings. Orange and blue retain
		the BH and MN colours of Fig.~\ref{fig:profiles}.
		Dashed, dotted, and solid curves correspond to
		$\lGB=-0.05,0,0.05$, respectively. The solid curves reproduce
		the corresponding profiles in Fig.~\ref{fig:profiles}.
		Insets magnify the indicated intervals. Each curve uses
		its own $u_h=n/(\pi T)$ in the horizontal coordinate.
		The infrared data are fixed within each panel, while the
		resulting UV masses generally differ.}
	\label{fig:coupling_profiles}
\end{figure}
\FloatBarrier

Figure~\ref{fig:profiles} illustrates these two ways of completing
the brane in the interior. Within each family, changing the regular
infrared datum generates a different UV mass and response.
Figure~\ref{fig:coupling_profiles} shows how this relation changes
with the GB coupling.

For these representative profiles, positive coupling lowers the
embedding angle relative to the Einstein solution at fixed $u/u_h$,
while negative coupling raises it. This ordering describes a change
in the map from infrared data to the UV source. Since the comparison
fixes $\theta_h$ or $u_0/u_h$, it does not establish which embedding
is preferred at a common physical mass. That question requires the
free-energy comparison below.

\subsection{Free-energy competition and the first-order transition}
\label{sec:thermo}

The embedding equation determines stationary configurations.
To select the equilibrium phase, we compare their renormalized
free-energy densities at identical sources. We follow probe-brane
holographic renormalization~\cite{KarchOBannonSkenderis2006}, with
the axial-source counterterms fixed as in the Einstein D3/D7
construction~\cite{Fadafan}. Denoting the physical
flavour free-energy density by $F$, we retain the normalized quantity
$\F=F/\Ncal$ introduced above:
\begin{equation}
	\F
	=
	\lim_{\epsilon\to0}
	\left[
	\int_\epsilon^{u_{\mathrm{end}}}
	du\,\mathcal I(u)
	+\mathrm{CT}(\epsilon)
	\right],
	\label{eq:free}
\end{equation}
where $u_{\mathrm{end}}=u_h$ for BH embeddings and
$u_{\mathrm{end}}=u_0$ for MN embeddings. Both families require the
same subtraction because their UV divergences are determined by the
same sources. We use
\begin{equation}
	\mathrm{CT}
	=
	-\frac{1}{4\epsilon^4}
	+\frac{\theta_\epsilon^2}{2\epsilon^4}
	-\frac{5\theta_\epsilon^4}{12\epsilon^4}
	+\frac{b^2\theta_\epsilon^2}{\epsilon^2}
	\left(
	\frac12\log|\theta_\epsilon|+\frac14
	\right).
	\label{eq:CT}
\end{equation}
Using Eq.~\eqref{eq:UV}, the on-shell radial density near the boundary
is $u^{-5}-M^2u^{-3}+b^2M^2u^{-1}+\cdots$. In particular,
$b^2u^2\sin^2\theta=b^2M^2u^4+\cdots$, so its square is
integrable and produces no separate $b^4\log\epsilon$ divergence.
The logarithm from expanding $\theta_\epsilon^2/(2\epsilon^4)$
and the explicit logarithm in Eq.~\eqref{eq:CT} together cancel the
logarithm of the regulated integral. Higher positive powers of
$\epsilon$ times logarithms vanish in the cutoff limit.
This prescription gives $\F(M=0)=-1/(4u_h^4)$ at finite
temperature. A common finite counterterm depending only on the
sources shifts both branches equally and leaves their crossing
unchanged. Further details of the subtraction are given in
Appendix~\ref{app:conventions}.

Following Ref.~\cite{Fadafan}, we express the physical mass response
as $\langle\mathcal O_m\rangle=\partial F/\partial m$. In the present
coordinates and renormalization scheme,
\begin{equation}
	\langle\mathcal O_m\rangle
	=
	\frac{\sqrt{\lambda_{\mathrm{tH}}}N_fN_c}{8\pi^3}
	\left[
	-2c_3+\frac{M^3}{3}
	+\frac{b^2M}{2}
	+b^2M\log\!\left(\frac{|M|}{\Lambda_{\rm UV}}\right)
	\right].
	\label{eq:massresponse}
\end{equation}
The same reference scale $\Lambda_{\rm UV}$ is used in
Eq.~\eqref{eq:UV}; our numerical units set $\Lambda_{\rm UV}=1$.
The expression at $M=0$ is understood by continuity.
For compactness in the analytical relations used later, we also define
the auxiliary normalized response
\begin{equation}
	Q
	\equiv
	\left.\frac{\partial\F}{\partial M}\right|_{T,b,\lGB}
	=
	\frac{8\pi^3}
	{\sqrt{\lambda_{\mathrm{tH}}}N_fN_c}
	\langle\mathcal O_m\rangle.
	\label{eq:Q}
\end{equation}
Since this operator is conjugate to an explicitly applied mass,
its discontinuity diagnoses the transition without implying
spontaneous symmetry breaking.

Define the differences between the two regular branches by
\begin{equation}
	\Delta\F=\F_{\BH}-\F_{\MN},
	\qquad
	\Delta Q=Q_{\BH}-Q_{\MN}.
	\label{eq:deltas}
\end{equation}
The coexistence mass is determined by
\begin{equation}
	\Delta\F(M_t,T,b,\lGB)=0,
	\qquad
	\left.\partial_M\Delta\F\right|_{M_t}=\Delta Q_t.
	\label{eq:coexistence}
\end{equation}
A negative $\Delta\F$ favours the BH branch, whereas a positive
$\Delta\F$ favours the MN branch among the compared homogeneous
solutions. Near a simple crossing,
\begin{equation}
	\Delta\F
	=
	\Delta Q_t(M-M_t)
	+\mathcal O\!\left((M-M_t)^2\right).
	\label{eq:localcrossing}
\end{equation}
For the positive $\Delta Q_t$ found here, the BH phase is preferred
below $M_t$ and the MN phase above it. The equilibrium free energy
is continuous at the crossing, but its mass derivative is
discontinuous.

\begin{figure}[htbp]
	\centering
	\manuscriptfigure[\textwidth]{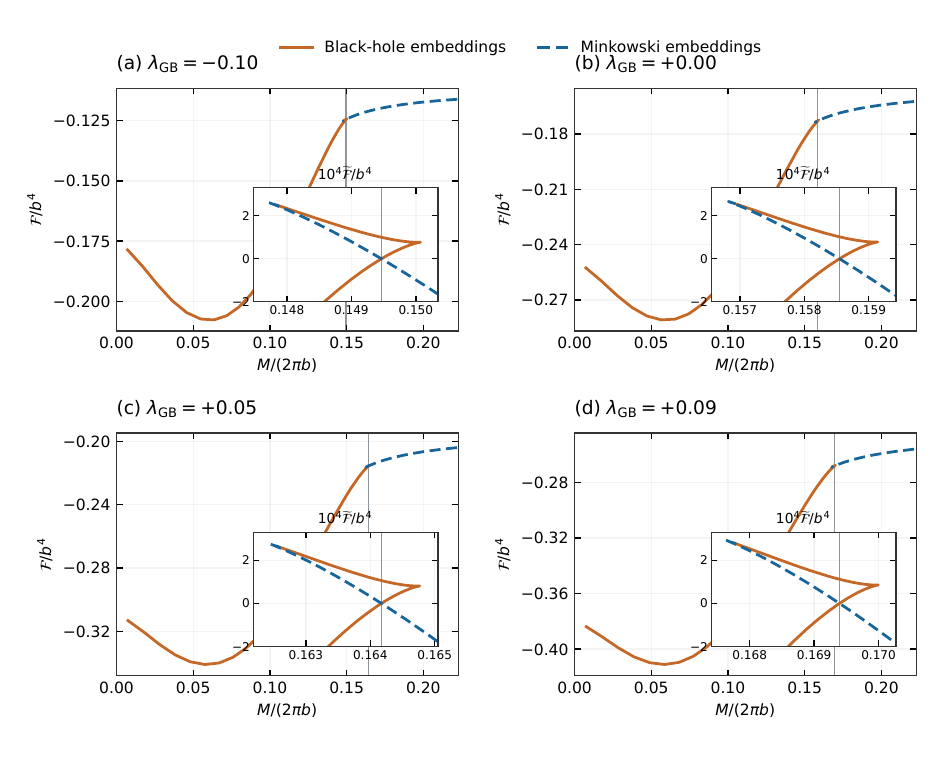}
	\caption{Free-energy competition at $T/b=1/\pi$ for
		$\lGB=-0.10,0,0.05,0.09$. The $-0.10$ panel illustrates the branch structure outside
        the three couplings used for the phase diagram. Solid orange and dashed blue curves
		denote BH and MN embeddings, respectively. Gray vertical
		lines mark equal-free-energy crossings. Insets resolve the
		multivalued region after subtracting the same affine function
		from both branches,
		$\widetilde\F=\F-\F_t-s_\lambda(M-M_t)$, with
		$s_\lambda=(Q_{\BH,t}+Q_{\MN,t})/2$.
		This subtraction preserves the crossing and the difference
		between branch slopes. The inset ordinate is
		$10^4\widetilde\F/b^4$. The horizontal coordinate is
		$M/(2\pi b)=m/(b\sqrt{\lambda_{\mathrm{tH}}})$.}
	\label{fig:free}
\end{figure}

The swallow tails in Fig.~\ref{fig:free} display several stationary
embeddings at the same mass. Equilibrium follows the branch with
the lowest free energy and switches between two distinct embeddings
at coexistence. The turning points of the families and the limiting
topology-changing embedding are therefore different from the
thermodynamic transition: neither determines $M_t$ without the
free-energy comparison.

The mass-response curves provide the complementary signature of
this same crossing. At coexistence the source-dependent terms in
Eq.~\eqref{eq:massresponse} agree on the two branches, giving
\begin{equation}
	\Delta Q_t
	=
	-2\left(c_{3,\BH}-c_{3,\MN}\right)_{M=M_t},
	\qquad
	\Delta\langle\mathcal O_m\rangle_t
	=
	\frac{\sqrt{\lambda_{\mathrm{tH}}}N_fN_c}{8\pi^3}
	\Delta Q_t.
	\label{eq:condensatejump}
\end{equation}
Although individual expectation values depend on the finite
renormalization scheme, their difference at common sources is
unchanged by a common finite counterterm. The unequal responses
therefore give the discontinuity in the slope of the equilibrium
free energy.

\begin{figure}[htbp]
	\centering
	\manuscriptfigure[\textwidth]{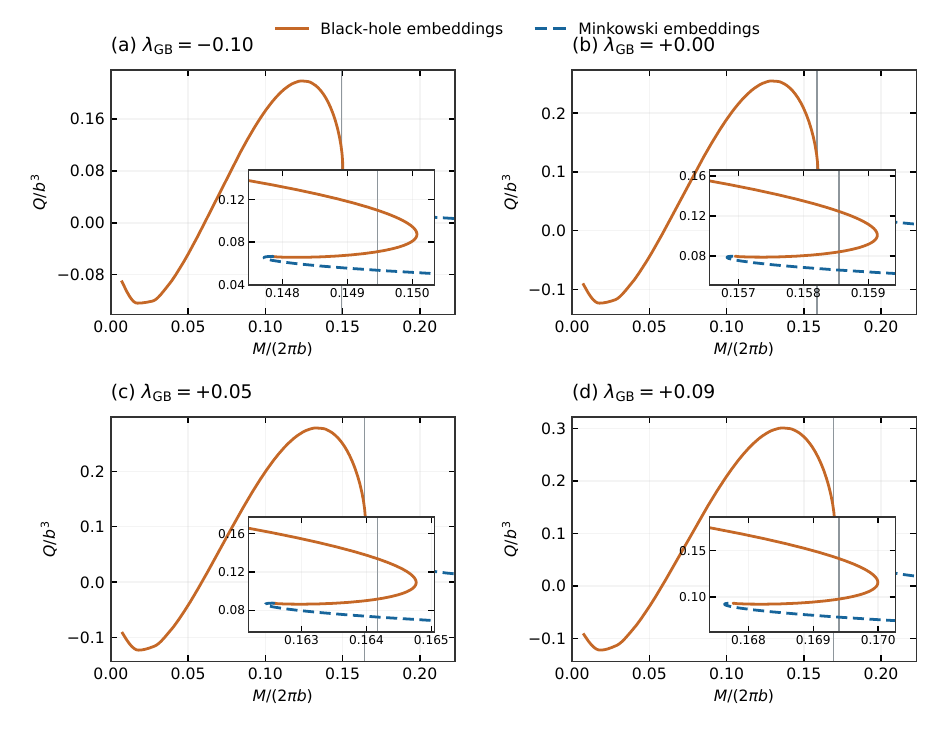}
	\caption{Mass response at $T/b=1/\pi$ for the couplings of
		Fig.~\ref{fig:free}. The normalized ordinate is
		$8\pi^3\langle\mathcal O_m\rangle/
		(\sqrt{\lambda_{\mathrm{tH}}}N_fN_c\,b^3)=Q/b^3$.
		Solid orange and dashed blue curves denote BH and MN
		embeddings. Gray vertical lines mark the free-energy
		crossings; the insets resolve the unequal expectation
		values at the same coexistence mass. Together with
		Fig.~\ref{fig:free}, these discontinuities establish the
		first-order transition. Multivalued portions outside the
		preferred pieces represent additional stationary solutions;
		their dynamical stability is not determined by these plots.}
	\label{fig:condensate_gb}
\end{figure}
\FloatBarrier

\subsection{Gauss--Bonnet phase boundary}
\label{sec:results}

\begin{figure}[htbp]
	\centering
	\manuscriptfigure[0.82\textwidth]{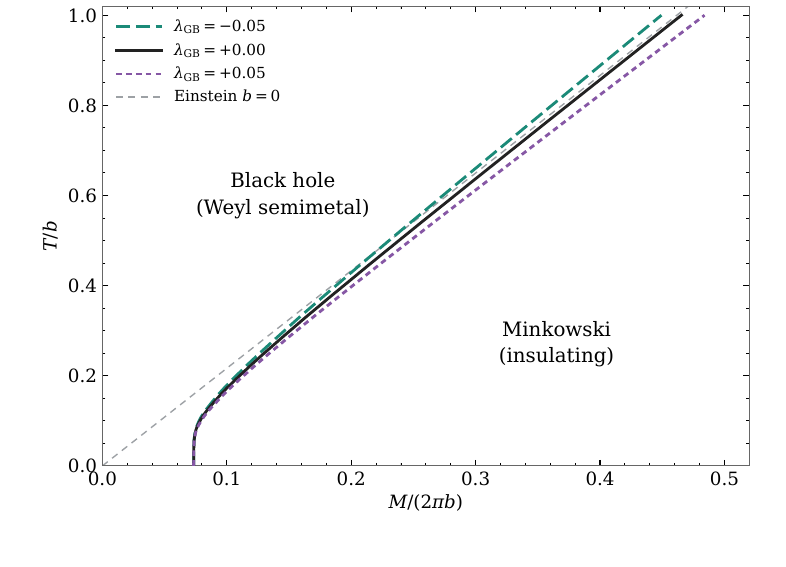}
	\caption{Mass--temperature phase diagram for
		$\lGB=-0.05$ (teal dashed), $0$ (black solid), and $0.05$
		(purple dotted). Each curve marks equal renormalized free
		energies of the BH and MN embeddings. The conducting BH phase
		is preferred to the left of a boundary and the insulating MN
		phase to its right. Positive GB coupling shifts the boundary
		toward larger mass at fixed physical temperature, extending
		the preferred BH region. The curves interpolate eleven
		directly computed finite-temperature crossings per coupling
		over $0.05\leq T/b\leq1$ and the common zero-temperature
		crossing,
		$m_t/(b\sqrt{\lambda_{\mathrm{tH}}})\simeq0.07333$.
		The segments below $T/b=0.05$ are interpolations.
		The thin gray line is the Einstein $b=0$ reference slope
		of Ref.~\cite{Fadafan}, expressed in the displayed variables.}
	\label{fig:phase_diagram}
\end{figure}

Having selected the lower-free-energy embedding at each common set of
physical sources, we obtain the phase boundaries in
Fig.~\ref{fig:phase_diagram}. Positive GB coupling shifts the
transition toward larger flavour mass at fixed physical temperature;
negative coupling produces the opposite trend in the sampled range.
The boundary is expressed using the mass convention in
Eq.~\eqref{eq:mass}, with $M_t$ the UV coexistence coefficient.

\FloatBarrier

At fixed temperature, increasing the mass eventually replaces the
horizon-reaching embedding by a capped embedding. Positive GB
coupling delays this change to a larger mass in the sampled
parameter range. Equivalently, on the monotonically rising parts
of the phase boundaries, increasing the coupling lowers the
temperature required to enter the BH phase at fixed mass.
These are two descriptions of the same displacement in
Fig.~\ref{fig:phase_diagram}.

Since the two coexisting branches have different mass responses
[Eq.~\eqref{eq:condensatejump}], this shifted boundary remains
first order at the sampled couplings. Its displacement is a change
in the competition between optimized free energies, rather than a
criterion read off from an individual embedding profile. A larger
conducting region need not imply larger conductivities, as we show
in Section~\ref{sec:transport}.

The ordering of the boundaries is supported at the sampled crossings;
the interpolation does not resolve the interval $0<T/b<0.05$ or
establish a critical endpoint. Numerical matching residuals and
selected refinement checks are reported in Appendices~\ref{app:conventions} and \ref{app:analyticchecks};
they are not uniform convergence bounds for the full phase diagram.

\subsection{Analytical displacement and temperature normalization}
\label{sec:responseidentity}

We can predict the leading boundary shift directly from the
coexistence condition~\eqref{eq:coexistence} and the response
\eqref{eq:Q}. Differentiating
$\Delta\F(M_t,T,b,\lGB)=0$ gives
\begin{equation}
	\left.\frac{dM_t}{d\lGB}\right|_{T,b}
	=
	-\frac{
		\left.\partial_{\lGB}\Delta\F\right|_{M,T,b}
	}{\Delta Q_t}.
	\label{eq:couplingslope}
\end{equation}
The derivative is taken at fixed physical temperature and axial
scale, with the retained flavour dictionary held fixed.
For a nonzero mass-response jump the crossing is simple, so this
derivative is well defined. When the jump is positive, lowering the BH free energy
relative to the MN free energy at the old crossing gives a
positive displacement of $M_t$.

An independent prediction follows by evaluating the explicit
variation of the probe action on the coexisting Einstein
embeddings. Define
\begin{equation}
	h_0=\frac{1}{\pi T},
	\qquad
	v=\left(\frac{u}{h_0}\right)^4,
	\qquad
	M_t^{(0)}=M_t(T,b,0).
\end{equation}
At fixed $(u,T)$, the blackening function and horizon position
satisfy
\begin{equation}
	f(u;\lGB,T)
	=
	1-v+\lGB(v^2-5v)+\mathcal O(\lGB^2),
	\qquad
	\left.\partial_{\lGB}u_h\right|_{T,0}=-h_0.
	\label{eq:fixedTvariation}
\end{equation}
For either coexisting Einstein embedding $\theta_0(u)$, introduce
\begin{equation}
	W(u)
	=
	\frac{
		\cos^3\theta_0
		\sqrt{1+b^2u^2\sin^2\theta_0}\,
		\theta_0'^2
	}{
		2u^3\sqrt{1+u^2(1-v)\theta_0'^2}
	},
	\qquad
	\Xi_i
	=
	\left.\partial_{\lGB}\F_i\right|_{M,T,b,0}.
	\label{eq:responseweight}
\end{equation}
Stationarity removes the bulk variation of the embedding at
fixed source. Including the displacement of the BH endpoint gives
\begin{align}
	\Xi_{\BH}
	&=
	\int_0^{h_0}du\,
	W_{\BH}(u)(v^2-5v)
	-
	\frac{
		\cos^3\theta_h
		\sqrt{1+b^2h_0^2\sin^2\theta_h}
	}{h_0^4},
	\nonumber\\
	\Xi_{\MN}
	&=
	\int_0^{u_0}du\,
	W_{\MN}(u)(v^2-5v).
	\label{eq:onshellresponse}
\end{align}
The horizon angle and cap position in these expressions belong to
the unperturbed embeddings. The integrals are UV finite. The smooth MN cap contributes no
endpoint term, while the UV contributions cancel against the
counterterm variation at fixed mass. 

The coexistence mass therefore obeys
\begin{equation}
	M_t
	=
	M_t^{(0)}
	+\lGB M_t^{(1)}
	+\mathcal O(\lGB^2),
	\qquad
	M_t^{(1)}
	=
	-\frac{\Xi_{\BH}-\Xi_{\MN}}{\Delta Q_0},
	\label{eq:perturbativeshift}
\end{equation}
where all quantities on the right are evaluated at Einstein
coexistence. The initial shift can thus be predicted from the
unperturbed embeddings without solving their first-order
deformations. Independently computed crossings at small positive
and negative coupling reproduce this prediction at
$T/b=0.10,1/\pi,0.50$, with a largest relative discrepancy of
$0.017\%$ across the tested settings. 

The physical-temperature convention is essential to interpreting
this response. Since $u_h=n/(\pi T)$, varying $\lGB$ at fixed $T$
also changes the horizon position. Treat $M_t$ as a function of
$(u_h,\lGB,b)$. Since $n'(0)=-1$, at the Einstein point
$(\partial_{\lGB}u_h)_{T}=-u_h$, while
$(\partial_Tu_h)_{\lGB=0}=-u_h/T$. The chain rule at unchanged
axial source $b$ therefore gives
\begin{equation}
	\left.\partial_{\lGB}M_t\right|_{T,b,0}
	=
	\left.\partial_{\lGB}M_t\right|_{u_h,b,0}
	+
	T\left.\frac{\partial M_t^{(0)}}{\partial T}\right|_b.
	\label{eq:slopedecomposition}
\end{equation}
The first term measures the change at fixed horizon position;
the second accounts for the change of thermal scale needed to
maintain the same physical temperature. At the three tested
temperatures, these contributions have opposite signs:
the fixed-horizon contribution is negative, while the positive
thermal contribution is larger, giving the positive
fixed-temperature slope.

The intrinsic term can also be expressed directly as a difference of
radial integrals over the coexisting Einstein embeddings. At fixed $u_h$,
with $v=(u/u_h)^4$ and $W$ defined in Eq.~\eqref{eq:responseweight},
\begin{equation}
 \left.\frac{\partial M_t}{\partial\lGB}\right|_{u_h,b,0}
 =\frac{1}{\Delta Q_0}\left[
 \int_0^{u_h}\!du\,W_{\BH}(u)\,v(1-v)
 -\int_0^{u_0}\!du\,W_{\MN}(u)\,v(1-v)
 \right].
 \label{eq:intrinsicfixedh}
\end{equation}
The fixed-horizon contribution is therefore a genuine bulk-shape
response, even within this effective prescription. The archived
Einstein-point coexistence checks in Table~\ref{tab:restoredslopes}
find fixed-horizon slopes $-0.026917$, $-0.199205$ and $-0.339305$
at $T/b=0.10$, $1/\pi$ and $0.50$, opposed by thermal contributions
$+0.129795$, $+0.852981$ and $+1.404065$, respectively. These are
first-order slopes at $\lGB=0$, not estimates of global monotonicity.
The corresponding analytical and direct finite-difference checks are tabulated in Appendix~\ref{app:analyticchecks}.

This decomposition connects the phase-boundary displacement to
the effective-temperature comparison discussed in
Section~\ref{sec:model}. An Einstein background at $T/n$ has the
same horizon position as the GB background at $T$. Part of the
observed displacement can therefore be understood through the
temperature dependence of the Einstein phase boundary.
The remaining difference reflects the deformation of the
blackening function at the matched horizon position.
Equation~\eqref{eq:slopedecomposition} makes this distinction
explicit at first order; it does not establish an exact
equivalence between the two geometries.

The common zero-temperature intercept of
Fig.~\ref{fig:phase_diagram} has a stronger, exact explanation.
At $T=0$, the normalized vacuum geometry satisfies
\begin{equation}
	u_h\longrightarrow\infty,
	\qquad
	f(u)=1.
\end{equation}
Because the internal metric, probe action, counterterms and
flavour dictionary are retained, the zero-temperature embedding
equations and normalized free-energy differences coincide with
those of the Einstein model. Consequently,
\begin{equation}
	\left.
	\frac{m_t}{b\sqrt{\lambda_{\mathrm{tH}}}}
	\right|_{T=0}
	=
	\left.\frac{M_t}{2\pi b}\right|_{T=0}
	\simeq0.07333.
\end{equation}
This coupling independence follows from the vacuum equations,
rather than from extrapolating the finite-temperature curves.
It is a property of the present geometric prescription:
a deformation that also changes the probe action can modify the
vacuum problem, as illustrated by the D-instanton construction
of Ref.~\cite{EomSeo2026}.

A complementary local check explains why the topology-changing
embedding retains its usual finite-temperature scaling.
Writing
\begin{equation}
	u=u_h(1-n\zeta^2),
	\qquad
	y=\frac{\pi}{2}-\theta,
	\label{eq:rindlercoords}
\end{equation}
the near-horizon geometry is Rindler and the shrinking
three-sphere contributes a factor $y^3$. The leading local action
and equation are
\begin{equation}
	I_{\mathrm{loc}}
	\propto
	\int d\zeta\,\zeta y^3
	\sqrt{1+(\partial_\zeta y)^2},
	\qquad
	\zeta y\,y''
	+(yy'-3\zeta)(1+y'^2)=0,
	\label{eq:criticalEOM}
\end{equation}
where derivatives are with respect to $\zeta$.
Linearizing about $y_c=\sqrt3\,\zeta$ gives
\begin{equation}
	\zeta^2\eta''+4\zeta\eta'+4\eta=0,
	\qquad
	\eta\propto\zeta^{\nu_\pm},
	\qquad
	\nu_\pm=-\frac32\pm\frac{i\sqrt7}{2}.
	\label{eq:criticalnu}
\end{equation}
These are the usual finite-temperature probe-brane
exponents~\cite{MateosThermo,Fadafan}.
The GB coupling changes the connection between the local
near-horizon region and the UV boundary, while leaving this
leading local equation unchanged. The exponents explain the
turning of nearby embedding families; they do not locate the
free-energy crossing or constitute critical exponents of the
equilibrium first-order transition. 

The phase boundary fixes the preferred embedding. We next use its
horizon data to determine the electrical response.

\section{Gauss--Bonnet corrections to DC conductivity}
\label{sec:transport}

The phase diagram determines where the conducting BH embedding is
thermodynamically preferred. We now examine how the GB deformation
changes its electrical response. Two results will be emphasized.
At the sampled low temperature, positive coupling reduces both
conductivity jumps at coexistence, despite extending the preferred
BH region. Near the temperature at which the longitudinal
conductivity becomes monotonic, it also suppresses and eventually
removes the mass-dependent maximum. The horizon formulas explain
these effects through the thermal scale and the selected embedding.

The conductivities
$\sigma_{xx}$ and $\sigma_{xy}$ describe the response in the
transverse $(x,y)$ plane. Our convention is
\begin{equation}
	J_x=\sigma_{xx}E_x+\sigma_{xy}E_y,
	\qquad
	J_y=-\sigma_{xy}E_x+\sigma_{xx}E_y,
	\label{eq:conductivitytensor}
\end{equation}
so an electric field along $x$ gives
$\sigma_{xx}=J_x/E_x$ and $\sigma_{xy}=-J_y/E_x$.
All conductivities refer to the probe flavour current and use the
same normalization
$K=N_fN_c/(4\pi^2)$.
The physical temperature relation
$u_h=n/(\pi T)$ is imposed separately at every coupling.

\subsection{Conductivity jumps and their GB dependence}
\label{sec:horizontransport}

The DC response follows from the radially conserved currents of
the quadratic DBI and WZ actions, using the standard probe-brane
current and horizon-regularity method~\cite{KarchOBannonMetallic2007,OBannonHall2007}. The Hall current
in our model is induced by the axial-source WZ term, rather than by
the external magnetic field used in some of those constructions.
Evaluating these currents with regular ingoing conditions at the
induced horizon gives
\begin{equation}
	\frac{\sigma_{xx}}{Kb}
	=
	\frac{\cos^3\theta_h}{bu_h}
	\sqrt{1+b^2u_h^2\sin^2\theta_h},
	\qquad
	\frac{\sigma_{xy}}{Kb}
	=
	\cos^4\theta_h.
	\label{eq:transport}
\end{equation}
These are the horizon expressions of the retained flavour-brane
construction~\cite{Fadafan}, evaluated in the prescribed GB geometry.
For a uniform DC field along $x$, the dissipative radial flux has
magnitude $K\cos^3\theta\sqrt{D/H}\,f|A_x'|/u$, with
$H=1+u^2f\theta'^2$. Regularity in ingoing coordinates gives
$f|A_x'|\to |E_x|$ and $H\to1$ at the horizon. Consequently, there is no additional
explicit factor of $n$ multiplying Eq.~\eqref{eq:transport}:
the coupling enters through $u_h$ and the regular horizon angle
$\theta_h$.

The factors in the two conductivities have distinct origins.
The wrapped-sphere volume contributes $\cos^3\theta_h$ to the
longitudinal response, while
$\sqrt{1+b^2u_h^2\sin^2\theta_h}$ accounts for the axial stretching
of the induced metric. The Hall response follows from the retained
WZ coupling and has the angular dependence $\cos^4\theta_h$.
Smooth MN embeddings have
$\sigma_{xx}=\sigma_{xy}=0$ in this zero-density probe prescription.

We evaluate these expressions on the embeddings selected by the
free-energy comparison. In the sampled region, the equilibrium
conductivities therefore satisfy
\begin{equation}
	\sigma_{ij}^{\mathrm{eq}}(m)
	=
	\begin{cases}
		\sigma_{ij}^{\BH}(m), & m<m_t,\\
		0, & m>m_t,
	\end{cases}
	\qquad ij=xx,xy.
	\label{eq:equilibriumconductivity}
\end{equation}
The discontinuities compare the two pure-phase limits:
\begin{equation}
	\Delta\sigma_{ij}
	\equiv
	\lim_{m\to m_t^-}\sigma_{ij}^{\mathrm{eq}}(m)
	-
	\lim_{m\to m_t^+}\sigma_{ij}^{\mathrm{eq}}(m)
	=
	\sigma_{ij}^{\BH}(m_t).
	\label{eq:transportjump}
\end{equation}
They are evaluated at each coupling's own coexistence mass.
They do not describe the effective response of a spatial mixture
at coexistence.

\begin{figure}[htbp]
	\centering
	\manuscriptfigure[0.99\linewidth]{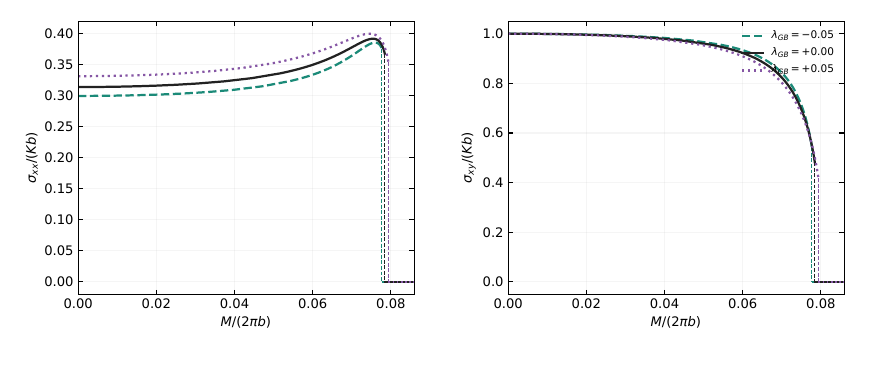}
	\caption{Equilibrium longitudinal and anomalous Hall DC
		conductivities at common physical temperature $T/b=0.10$
		for $\lGB=-0.05$ (teal dashed), $0$ (black solid), and
		$0.05$ (purple dotted). Both conductivities are normalized
		by $Kb$, and the mass coordinate is
		$m/(b\sqrt{\lambda_{\mathrm{tH}}})=M/(2\pi b)$.
		The finite curves follow the preferred BH embeddings up
		to their respective free-energy crossings. Vertical dashed
		segments mark the jumps to the vanishing MN conductivities.
		Positive coupling moves coexistence to larger mass while
		reducing both terminating conductivity values.
		The vertical segments indicate discontinuities between
		homogeneous phases.}
	\label{fig:transport}
\end{figure}
\FloatBarrier

Figure~\ref{fig:transport} shows that both jumps decrease as
$\lGB$ increases over the sampled interval at $T/b=0.10$.
Between $\lGB=-0.05$ and $0.05$, the coexistence mass increases
by approximately $2.13\%$, whereas the longitudinal and Hall
jumps decrease by approximately $5.77\%$ and $22.85\%$,
respectively. Thus the Hall discontinuity is more sensitive
over this interval. Negative coupling gives larger jumps than
the Einstein result, while positive coupling gives smaller ones.
This ordering is established at $T/b=0.10$ only. In particular,
the sign reverses for the Ohmic jump at lower sampled temperatures,
as shown below.

The massless embedding provides a useful exact reference at $T>0$.
For $\theta_h=0$, Eq.~\eqref{eq:transport} becomes
\begin{equation}
	\frac{\sigma_{xx}(0)}{K}
	=
	\frac{1}{u_h}
	=
	\frac{\pi T}{n},
	\qquad
	\frac{\sigma_{xy}(0)}{Kb}=1.
	\label{eq:masslesstransport}
\end{equation}
Positive coupling therefore increases the massless longitudinal
conductivity at fixed physical temperature, while leaving the
massless Hall value unchanged. The decrease of the conductivity
jumps must consequently involve the changing embedding at
coexistence.

This interpretation can be made precise without assuming the
sign of the embedding response. Let
$\theta_{h,t}=\theta_h(M_t(T,b,\lGB),T,b,\lGB)$ and define
$D_t=1+b^2u_h^2\sin^2\theta_{h,t}$.
Differentiating Eq.~\eqref{eq:transport} along the coexistence
line at fixed physical $(T,b)$ gives
\begin{align}
	\frac{d}{d\lGB}
	\ln\!\left(\frac{\Delta\sigma_{xx}}{Kb}\right)
	&=
	-\frac{1}{D_t}\frac{d\ln n}{d\lGB}
	+
	\left[
	-3\tan\theta_{h,t}
	+\frac{
		b^2u_h^2\sin\theta_{h,t}\cos\theta_{h,t}
	}{D_t}
	\right]
	\frac{d\theta_{h,t}}{d\lGB},
	\label{eq:ohmicsensitivity}\\
	\frac{d}{d\lGB}
	\ln\!\left(\frac{\Delta\sigma_{xy}}{Kb}\right)
	&=
	-4\tan\theta_{h,t}
	\frac{d\theta_{h,t}}{d\lGB}.
	\label{eq:hallsensitivity}
\end{align}
These identities are exact consequences of the horizon formulas.
Their interpretation requires the coexistence embeddings obtained
numerically.

For a compact sign test define
\begin{equation}
 G_t=3\tan\theta_{h,t}-\frac{b^2u_h^2\sin\theta_{h,t}
 \cos\theta_{h,t}}{D_t}.
\end{equation}
When $G_t>0$, suppression of the Ohmic jump requires
\begin{equation}
 G_t\,\frac{d\theta_{h,t}}{d\lGB}>
 -\frac{1}{D_t}\frac{d\ln n}{d\lGB}.
 \label{eq:ohmicsigncriterion}
\end{equation}
The horizon-angle derivative is a total derivative along coexistence:
\begin{equation}
 \left.\frac{d\theta_{h,t}}{d\lGB}\right|_{T,b}
 =\left.\frac{\partial\theta_h}{\partial\lGB}\right|_{M,T,b}
 +\left.\frac{\partial\theta_h}{\partial M}\right|_{T,b,\lGB}
 \left.\frac{dM_t}{d\lGB}\right|_{T,b}.
 \label{eq:thetahsensitivitysplit}
\end{equation}
Only the sum in Eq.~\eqref{eq:thetahsensitivitysplit} is evaluated
in the present numerical table; we do not claim separately converged
values for its two terms.

\begin{table}[htbp]
\centering\small
\caption{Longitudinal DC conductivity jump along coexistence,
$\Delta\sigma_{xx}/(Kb)$, at several physical temperatures.
The $T/b=0.070$ and $0.085$ crossings have been re-solved using
matched UV masses and directly renormalized BH/MN free energies;
other rows are from the archived coexistence dataset. Values are
BH-side limits, not conductivities at a common mass.}
\label{tab:conductivitytemperature}
\begin{tabular}{@{}rccc@{}}\hline
$T/b$ & $\lGB=-0.05$ & $\lGB=0$ & $\lGB=+0.05$\\\hline
0.050 & 0.150854 & 0.158606 & 0.168464\\
0.070 & 0.242838 & 0.264987 & 0.293493\\
0.075 & 0.280121 & 0.305729 & 0.334463\\
0.085 & 0.351400 & 0.367867 & 0.378432\\
0.100 & 0.375739 & 0.367854 & 0.354067\\
0.150 & 0.238217 & 0.223511 & 0.208212\\\hline
\end{tabular}
\end{table}
The increase at $T/b=0.070$ confirms that suppression is not
universal over the sampled coupling interval. The finite-interval
ordering of the three couplings reverses between $T/b=0.085$ and
$0.10$; the six-temperature table does not locate the precise
crossover or establish a sign change of the infinitesimal derivative
at every coupling. A finer scan is needed for those stronger claims.

Since
$d\ln n/d\lGB=-1/[n\sqrt{1-4\lGB}]<0$,
the explicit thermal-scale contribution in
Eq.~\eqref{eq:ohmicsensitivity} is positive.
At the sampled crossings, $\theta_{h,t}$ increases with
coupling, which suppresses the Hall jump. The angular part of
the Ohmic identity can have either sign: its competition with
the positive explicit thermal-scale term changes with temperature.
The identity therefore organizes the computed response but does not
by itself predict suppression.

The angular change includes both the deformation of the BH
solution at fixed mass and the displacement of the coexistence
mass itself. The identities therefore explain the numerical
trend but do not independently establish its sign.
Likewise, the comparison with Einstein gravity at $T/n$ can
account for part of this response through its effect on the
coexistence embedding; the horizon formulas alone do not
quantify the remaining GB contribution.

The central physical distinction is between the extent of the
conducting phase and its response at the boundary of that phase.
The former follows from a difference of optimized free energies,
while the latter depends on local horizon data. Positive coupling
can consequently extend the preferred BH region while decreasing
or increasing the Ohmic jump, depending on temperature. With the WZ normalization fixed, this
finite-temperature Hall suppression does not represent a
correction to the axial-anomaly coefficient.

\subsection{Matched-horizon comparison and residual GB response}
\label{sec:shapecheck}
To quantify the part of the response not explained by the thermal
scale alone, we compare GB coexistence at physical $T$ with
Einstein coexistence at $T_{\rm eff}=T/n(\lGB)$, so that $u_h$
is the same. Both points lie on their own coexistence lines:
this is not a comparison at equal flavour mass. For an observable
$\mathcal O$ define the relative residual
\begin{equation}
 \delta_{\rm shape}(\mathcal O)=
 \frac{\mathcal O_{\rm GB}(T,b,M_t(\lGB))}
 {\mathcal O_{\rm E}(T/n,b,M_t(0;T/n))}-1.
 \label{eq:shaperesidual}
\end{equation}
The residual combines the explicit difference of blackening functions
and the resulting shift of the optimized embedding. It does not
isolate a unique local higher-curvature operator.
\begin{table}[htbp]
\centering\small
\caption{Percentage residual $100\delta_{\rm shape}$ at $T/b=0.10$,
from the archived matched-horizon BH-side coexistence solutions.
The two conductivity jumps use the retained current
normalization $K$; each theory is evaluated on its own coexistence line.}
\label{tab:matchedshape}
\begin{tabular}{@{}lrr@{}}\hline
Observable & $\lGB=-0.05$ & $\lGB=+0.05$\\\hline
$M_t/b$ & +0.210\% & -0.365\%\\
$\Delta\sigma_{xx}/(Kb)$ & -0.257\% & +0.603\%\\
$\Delta\sigma_{xy}/(Kb)$ & -1.142\% & +1.499\%\\

\hline
\end{tabular}
\end{table}
The residuals in Table~\ref{tab:matchedshape} are small on these
coexistence states (at most $1.5\%$ in magnitude for the listed
observables). Thus thermal rescaling reproduces most of the response
for the listed quantities at $T/b=0.10$, but not exactly. These
percentages quantify differences between two coexistence solutions,
not the fraction of a coupling derivative due to the intrinsic term.
For the latter, the analytical phase-boundary decomposition in
Eq.~\eqref{eq:intrinsicfixedh} gives a more direct interpretation.
The available matched-horizon data do not establish a similar bound
at other physical temperatures or at fixed mass.

\subsection{Suppression and disappearance of the longitudinal maximum}
\label{sec:dcpeak}

The longitudinal response contains a second feature beyond its
discontinuity at coexistence. The original Einstein model exhibits
a maximum of the DC conductivity as the mass is varied at
sufficiently low temperature, followed by monotonic behavior at
higher temperature~\cite{Fadafan}. The D-instanton extension also
reports suppression of this low-temperature maximum when the
instanton density is introduced~\cite{EomSeo2026}.
The disappearance of such a maximum is therefore not unique
to GB corrections. Here the retained horizon formula allows
its geometric condition to be determined analytically and
tested on the preferred embeddings.

At fixed $(T,b,\lGB)$, introduce
$B=bu_h=bn/(\pi T)$ and $x=\sin^2\theta_h$.
Dividing the longitudinal conductivity by its massless value
at the same temperature and coupling gives
\begin{equation}
	\mathcal R(x)
	\equiv
	\frac{\sigma_{xx}(\theta_h)}{\sigma_{xx}(0)}
	=
	(1-x)^{3/2}\sqrt{1+B^2x}.
	\label{eq:peakratio}
\end{equation}
The first factor decreases as the wrapped sphere shrinks,
whereas the second increases because of axial stretching.
Their competition is expressed by
\begin{equation}
	\frac{\partial\ln\mathcal R}{\partial x}
	=
	\frac{B^2-3-4B^2x}
	{2(1-x)(1+B^2x)}.
	\label{eq:peakderivative}
\end{equation}
For $B\leq\sqrt3$, the conductivity decreases monotonically
with the horizon angle. For $B>\sqrt3$, there is exactly one
interior maximum:
\begin{equation}
	\sin^2\theta_h^*
	=
	\frac14\left(1-\frac{3}{B^2}\right),
	\qquad
	\mathcal R_{\max}
	=
	\frac{3\sqrt3(1+B^2)^2}{16B^3}.
	\label{eq:peakmax}
\end{equation}
The geometric existence condition is therefore
\begin{equation}
	\frac{T}{b}
	<
	\left(\frac{T}{b}\right)_{\mathrm{geom}}^*
	\equiv
	\frac{n(\lGB)}{\pi\sqrt3}.
	\label{eq:peakthreshold}
\end{equation}

Positive coupling decreases $n$ and lowers this threshold.
It also reduces the maximum relative to the massless reference
while $B>\sqrt3$. In terms of
$T_{\mathrm{eff}}=T/n$, Eq.~\eqref{eq:peakthreshold} is precisely
the Einstein condition
$T_{\mathrm{eff}}/b<1/(\pi\sqrt3)$.
The GB dependence of this geometric criterion thus follows
directly from the thermal-scale relation.

An interior maximum of the horizon formula need not appear
in the equilibrium conductivity against mass. The preferred
BH family must reach $\theta_h^*$ before its free-energy
crossing. Moreover, the local source map must be regular,
$\partial M/\partial\theta_h\ne0$, so the angular extremum
corresponds to an extremum against mass. A monotonic map along
the preferred branch makes this correspondence unambiguous.
Phase selection can therefore exclude the maximum even when
Eq.~\eqref{eq:peakthreshold} is satisfied.

We test this distinction at $T/b=0.18$, where the threshold
passes through the sampled coupling range.
The preferred BH branches have a monotonic mass--angle map
up to coexistence. For $\lGB=-0.05$ and $0$, the maximum is
reached before the first-order transition, with relative
enhancements of approximately $0.333\%$ and $0.032\%$ above
each coupling's own massless conductivity. For $\lGB=0.05$,
the geometric threshold is $T/b\simeq0.17408$, below the
chosen temperature, and the conductivity decreases
monotonically with mass.

\begin{figure}[htbp]
	\centering
	\manuscriptfigure[.8\textwidth]{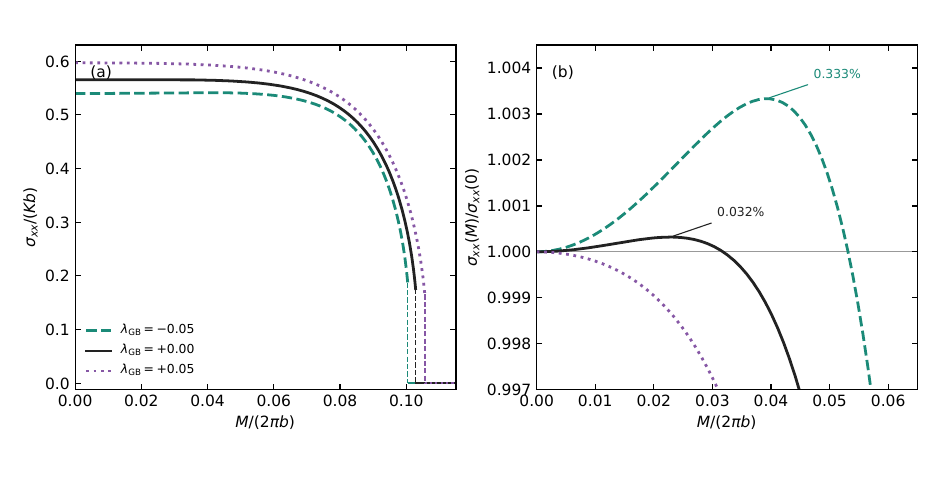}
	\caption{Suppression and disappearance of the equilibrium
		longitudinal DC maximum at common physical temperature
		$T/b=0.18$. Teal dashed, black solid, and purple dotted
		curves correspond to $\lGB=-0.05,0,0.05$, respectively.
		Panel (a) follows the preferred BH embeddings to their
		free-energy crossings, where the conductivity jumps to
		the vanishing MN value. Panel (b) enlarges the low-mass
		region after division by each coupling's own massless
		conductivity. The negative-coupling and Einstein curves
		have shallow maxima, while the positive-coupling curve
		is monotonic. The percentages denote the relative
		enhancements above the corresponding massless values.
		The mass coordinate is
		$m/(b\sqrt{\lambda_{\mathrm{tH}}})=M/(2\pi b)$.}
	\label{fig:equilibriumpeak}
\end{figure}
\FloatBarrier

The enlarged panel in Fig.~\ref{fig:equilibriumpeak} resolves the
shallow Einstein maximum. Its small height follows from the
analytical behavior near the threshold. Writing
$\epsilon=1-T/T_{\mathrm{geom}}^*$ at fixed coupling gives
\begin{equation}
	\sin^2\theta_h^*
	=
	\frac{\epsilon}{2}+\mathcal O(\epsilon^2),
	\qquad
	\mathcal R_{\max}-1
	=
	\frac34\epsilon^2+\mathcal O(\epsilon^3).
	\label{eq:peakmerging}
\end{equation}
The relative enhancement therefore vanishes quadratically
as the maximum approaches the massless endpoint.
At fixed $T/b=0.18$, the geometric threshold corresponds to
$\lGB^*\simeq0.02013$. The same expansion applies when $\lGB$ is varied at fixed $T$,
since $\epsilon=1-\pi\sqrt{3}T/(b n)$ measures distance from the
threshold in either scan. At $\lGB^*-0.002$ the interior peak lies at
$M_*/b=0.04670$, with $\mathcal R_{\max}-1=3.38\times10^{-6}$,
whereas $M_t/b=0.65202$. At $\lGB^*+0.002$ there is no interior peak.
These embedding and free-energy checks support a smooth disappearance
through zero mass while coexistence remains at finite mass.
In this region the maximum is lost through the shape of the
conductivity curve, rather than being bypassed by the
first-order transition.

This conclusion concerns the DC conductivity as a function
of mass. It does not identify the appearance or disappearance
of a pole in the finite-frequency current correlator.
The optical structures and long-lived current modes studied
in the Einstein model~\cite{Furukawa2025} require a separate
GB fluctuation calculation.

The two transport results thus probe different aspects of
the same equilibrium family. The conductivity jumps compare
the horizon data selected at coexistence, whereas the
interior maximum measures the competition between axial
stretching and the shrinking wrapped sphere along the BH
branch. Their coupling dependence is constrained by exact
horizon relations, with the physical realization determined
by regular embeddings and free-energy selection.

\section{Conclusions and outlook}
\label{sec:conclusions}

We have examined how an effective Gauss--Bonnet deformation changes
the phase structure and DC transport of the D3/D7 holographic Weyl
semimetal. The background is an effective five-dimensional GB deformation
combined with the unchanged internal geometry and probe DBI/WZ
prescription. We do not claim a corrected ten-dimensional type-IIB
solution or systematic finite-'t Hooft-coupling expansion, and the
probe approximation does not assess all possible inhomogeneous or
dynamical instabilities. Comparing phases at common physical sources leads to the
following results:
\begin{itemize}
 \item \textbf{The first-order boundary moves while its character persists.}
 Positive GB coupling extends the preferred conducting BH phase
 toward larger flavour mass at finite temperature in the sampled
 range; negative coupling produces the opposite shift. The unequal
 mass responses at coexistence establish the first-order character
 of the transition. The normalized zero-temperature boundary remains
 unchanged because the vacuum probe problem is unchanged.

 \item \textbf{The leading boundary shift follows from the Einstein embeddings.}
 An on-shell action variation predicts the initial displacement
 without solving the first-order embedding deformation. Its agreement
 with independently computed crossings supports the interpretation
 of the shift as a competition between the two optimized free energies.
 The decomposition into fixed-horizon and thermal-scale contributions
 identifies the origin of the positive shift at the tested temperatures.

 \item \textbf{An enlarged conducting region need not have larger conductivity jumps.}
 At $T/b=0.10$, increasing the coupling suppresses both longitudinal
 and anomalous Hall discontinuities at their respective coexistence
 masses. The changing horizon embedding reduces the wrapped-sphere
 factors in the conductivities. For the longitudinal jump, this
 suppression outweighs the increase associated explicitly with the
 thermal scale. The Hall change follows from the embedding with the
 WZ normalization held fixed.

 \item \textbf{The longitudinal maximum can disappear within the preferred phase.}
 The horizon formula admits an interior maximum when
 $T/b<n(\lGB)/(\pi\sqrt3)$. This geometric criterion is the Einstein
 condition expressed at the matched-horizon temperature. Free-energy
 selection supplies the additional requirement that the preferred BH
 branch reaches the maximum. The targeted scan resolves its smooth
 merger with the massless endpoint, with a quadratically vanishing
 relative enhancement, while coexistence remains at finite mass.
\end{itemize}

The conductivity-jump suppression is not universal: at
$T/b=0.070$ the Ohmic jump increases over the same coupling interval.
The matched-horizon residuals in Table~\ref{tab:matchedshape} show
that thermal rescaling accounts for most of the listed responses at
$T/b=0.10$, while nonzero residuals test deformation-dependent
embedding and blackening-function effects. The intrinsic coexistence
slope is nonzero and of opposite sign to the thermal contribution at
the three temperatures checked analytically.

An important extension is to determine how the GB deformation
modifies the frequency- and momentum-dependent flavour-current
response. Optical conductivities, spectral functions and quasinormal
modes in the Einstein D3/D7 Weyl semimetal provide a direct baseline
for this question~\cite{Furukawa2025}. Tracking resonance positions,
widths and spectral weight on the thermodynamically preferred BH and
MN branches would test whether the GB background produces changes
beyond those reproduced by the matched-horizon effective temperature.
It would also clarify whether the discontinuities in DC transport
across coexistence have distinctive finite-frequency signatures.

Fermionic flavour excitations offer a complementary probe of the same
transition. The quadratic D7-brane fermion equations, with their
induced geometry, axial embedding and background Ramond--Ramond flux,
would need to be derived and solved on the competing embeddings.
Polarization-resolved retarded fermionic correlators could then be
used to follow spectral weight and pole structure across the
insulating--Weyl-semimetal transition. A useful objective would be to
identify any GB-dependent features that survive a matched-horizon
comparison and go beyond the expected distinction between normal
modes on capped embeddings and dissipative excitations on
horizon-reaching embeddings. These questions lie outside the present
bosonic embedding and DC-conductivity analysis.

\section*{Acknowledgements}
\addcontentsline{toc}{section}{Acknowledgements}
We thank Ronnie Rodgers for helpful discussions.
The initial embedding and free-energy analysis builds on
the code developed by the authors. The authors used OpenAI's ChatGPT to assist with analytical
derivations, additional numerical calculations and consistency checks, and manuscript editing.

\appendix

\section{Conventions and numerical checks}
\label{app:conventions}

\begin{itemize}
\item \textbf{Radial coordinates and physical temperature.}

The temperature relation depends on the radial convention,
although the physical Hawking temperature is invariant.
In the commonly used bare-radius convention~\cite{Brigante},
we denote the radial coordinate by $\bar u$ to distinguish it from
the boundary direction $z$. The metric is
\begin{equation}
	ds_5^2
	=
	\frac{L_0^2}{\bar u^2}
	\left[
	-n f_0(\bar u)\,dt^2+d\boldsymbol{x}^2
	+\frac{d\bar u^2}{f_0(\bar u)}
	\right],
	\qquad
	T=\frac{\sqrt n}{\pi \bar u_h}.
\end{equation}
Here
\begin{equation}
 f_0(\bar u)=\frac{1-\sqrt{1-4\lGB[1-(\bar u/\bar u_h)^4]}}{2\lGB},
\end{equation}
with the Einstein limit understood at $\lGB=0$.
Using $L_{\mathrm{eff}}^2=nL_0^2=1$ and defining
$u=\sqrt n\,\bar u$ gives the metric of Eq.~\eqref{eq:metric},
with $f=nf_0$ and $u_h=\sqrt n\,\bar u_h$. Consequently,
\begin{equation}
	T=\frac{\sqrt n}{\pi \bar u_h}
	=\frac{n}{\pi u_h}.
\end{equation}
The two expressions therefore describe the same temperature.
The horizon coordinates $\bar u_h$ and $u_h$ must not be interchanged
when comparing different couplings.

\item \textbf{Dimensionless shooting variables and the mass response.}

For finite-temperature shooting, we use $x=u/u_h$ and
$\widehat b=bu_h$. Define the dimensionless UV fit by
\begin{equation}
 \theta(x)=\widehat M x+
 \left[\widehat c_3+\frac{\widehat b^{\,2}\widehat M}{2}\log x\right]x^3+\cdots.
\end{equation}
Comparing with Eq.~\eqref{eq:UV} gives
\begin{equation}
	M=\frac{\widehat M}{u_h},
	\qquad
	c_3=
	\frac{
		\widehat c_3
		-\tfrac12\widehat b^{\,2}\widehat M\log(u_h\Lambda_{\rm UV})
	}{u_h^3},
	\qquad
	Q=\frac{\widehat Q}{u_h^3},
	\qquad
	\F=\frac{\widehat\F}{u_h^4}.
	\label{eq:uvconversion}
\end{equation}
Here $\widehat Q$ and $\widehat\F$ use the same subtraction as
Eqs.~\eqref{eq:CT} and \eqref{eq:Q}, with dimensionless reference
scale equal to one. The factors of $u_h\Lambda_{\rm UV}$ cancel in $Q$;
Eq.~\eqref{eq:uvconversion} reduces to the numerical convention
$\Lambda_{\rm UV}=1$ used in the data. The logarithmic conversion of $c_3$ is essential:
omitting it changes the extracted mass response.

In the Cartesian embedding notation of Ref.~\cite{Fadafan},
the corresponding subleading coefficient is
\begin{equation}
	C=c_3-\frac{M^3}{6},
\end{equation}
with the same logarithmic reference convention.
The physical expectation value can therefore be written
equivalently as
\begin{equation}
	\langle\mathcal O_m\rangle
	=
	\frac{\sqrt{\lambda_{\mathrm{tH}}}N_fN_c}{8\pi^3}
	\left[
	-2C+\frac{b^2M}{2}
	+b^2M\log\!\left(\frac{|M|}{\Lambda_{\rm UV}}\right)
	\right].
\end{equation}
This reproduces the notation of the original construction
while distinguishing $C$ from the coefficient $c_3$ of the
angular embedding.

Changing the reference scale at fixed embedding gives
\begin{equation}
	c_3(\Lambda_{\rm UV}')
	=
	c_3(\Lambda_{\rm UV})
	-\frac{b^2M}{2}
	\log\!\left(\frac{\Lambda_{\rm UV}'}{\Lambda_{\rm UV}}\right).
	\label{eq:scalereference}
\end{equation}
This change cancels the explicit scale dependence in
Eq.~\eqref{eq:massresponse}. It is a reparametrization of the
same UV solution, distinct from adding a finite counterterm
to change the renormalization scheme.

\item \textbf{Initialization and free-energy checks.}
For BH embeddings, the near-horizon expansion uses
Eq.~\eqref{eq:horizon}. For MN embeddings, the inverse-coordinate
series in Eq.~\eqref{eq:cap} is initialized near the smooth cap
and converted to the $u$ coordinate in an overlap region.
The ODE solver is DOP853. The UV fit includes the logarithm
in Eq.~\eqref{eq:UV}.

Solutions are continued in their infrared parameter rather
than sorted only by mass. This retains folds and multivalued
portions of the embedding families. Free energies are compared
only after matching the physical UV mass, using the same
$(T,b,\lGB)$ and subtraction prescription.

The leading $u^{-5}$ divergence is subtracted analytically
before numerical integration. The renormalized actions are
evaluated at five UV cutoffs and extrapolated using the basis
\begin{equation}
	1,\qquad
	\epsilon^2\log^2\epsilon,\qquad
	\epsilon^2\log\epsilon,\qquad
	\epsilon^2.
\end{equation}
As an additional check, direct action differences are compared
with differences obtained by integrating
$Q\,dM$ along each continued branch, with integration constants
fixed by direct action values, where $Q$ is the auxiliary normalized mass response
defined in Eq.~\eqref{eq:Q}.

For the four coexistence points at $T/b=1/\pi$, refinement
reduced the UV windows and cutoffs by a factor of $0.75$,
the infrared offsets by a factor of $0.20$, and the relative
ODE tolerance from $2\times10^{-12}$ to $5\times10^{-13}$.
The largest reported change in $M_t/b$ was
$1.3\times10^{-7}$. Matched-mass action comparisons on either
side of coexistence checked the sign change of $\Delta\F$.
A further refinement of the transport coexistence points at
$T/b=0.10$ changed the mass by less than $1.7\times10^{-8}$.

For the 33 finite-temperature phase-boundary pairs,
mass-matching residuals were below $1.08\times10^{-7}$ and
direct free-energy residuals below $9.4\times10^{-9}$.
These residuals test the matching equations; they are not
uniform convergence error bars for the entire phase diagram.
The sensitivity tests support the reported results within
the sampled homogeneous families and do not determine
fluctuation stability.

The flat-boundary subtraction used here and the subtraction
using the induced boundary volume element differ by
\begin{equation}
	\F_{\mathrm{cov}}-\F_{\mathrm{flat}}
	=
	\frac{n}
	{8\sqrt{1-4\lGB}\,u_h^4}.
	\label{eq:subtractionoffset}
\end{equation}
At common sources and coupling, this offset is identical
for BH and MN embeddings and cancels in $\Delta\F$.
Its temperature dependence must nevertheless be retained
when comparing absolute free energies or their thermal
derivatives. Because it contains $u_h^{-4}$, it should be
distinguished from a finite local counterterm depending
only on the fixed boundary sources.

\item \textbf{Resolution of the shallow conductivity maximum.}
\label{app:peakscan}

The peak analysis includes eighteen coexistence calculations
for $\lGB=-0.05,0,0.05$, covering
$T/b=0.17,0.18,0.19,0.20$ and temperatures $0.0005$ below
and above each geometric threshold. The map $M(\theta_h)$
was sampled below coexistence and was monotonic in these
tests.

Six points were repeated with the refined settings described
above. The largest change in $M_t/b$ was $6.3\times10^{-8}$.
The baseline mass and normalized free-energy matching
residuals were below $7.6\times10^{-10}$ and
$8.9\times10^{-11}$, respectively.

At $T/b=0.18$, broader scans of both embedding families were
followed by direct matched-mass action comparisons.
None of the sampled competitors had lower free energy
than the selected BH branch. At the two nonzero peak masses,
the broad scan found one matching BH solution and no MN
solution. These checks support the equilibrium interpretation
within the sampled families without proving that all stationary solutions have been found
or establishing dynamical stability.

Additional coexistence calculations were performed at
$\lGB^*$ and $\lGB^*\pm0.002$, where $\lGB^*$ denotes the
geometric threshold at $T/b=0.18$.
The peak heights are evaluated from the exact horizon formula
at $\theta_h^*$; they are not extracted from differences
between nearly equal free energies.

The two jumps also obey an algebraic relation because they
depend on the same horizon angle. Defining
\begin{equation}
	X=\frac{\Delta\sigma_{xx}}{Kb},
	\qquad
	Y=\frac{\Delta\sigma_{xy}}{Kb},
\end{equation}
eliminating $\theta_{h,t}$ gives
\begin{equation}
	X=
	\frac{Y^{3/4}}{bu_h}
	\sqrt{
		1+(bu_h)^2\left(1-\sqrt Y\right)
	},
	\qquad
	u_h=\frac{n}{\pi T}.
	\label{eq:jumpconsistency}
\end{equation}
This is an exact consistency relation within the retained
DBI/WZ prescription when the MN conductivities vanish.
It checks the common normalization and implementation,
with a maximum absolute residual in $X$ of $3.1\times10^{-16}$
over the 33 stored finite-temperature coexistence states. Both sides
use the same horizon data.

The angular dependence used in this check and the threshold of the DC maximum are shown in Fig.~\ref{fig:peakgeometry}.
\begin{figure}[htbp]
	\centering
	\manuscriptfigure[0.88\linewidth]{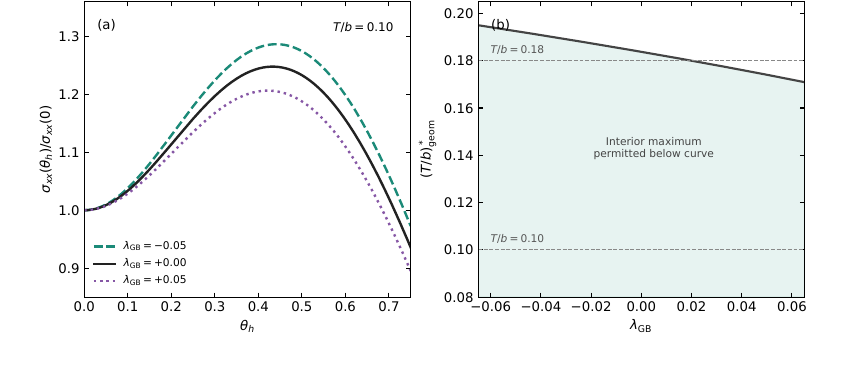}
	\caption{Geometric condition for a longitudinal DC maximum.
		Panel (a) shows the exact ratio
		$\mathcal R(\theta_h)=\sigma_{xx}(\theta_h)/\sigma_{xx}(0)$
		at $T/b=0.10$ for $\lGB=-0.05,0,0.05$.
		These are angle-parametrized horizon formulas,
		rather than equilibrium scans against mass.
		Panel (b) shows the threshold
		$(T/b)_{\mathrm{geom}}^*=n/(\pi\sqrt3)$.
		The shaded region permits an interior maximum.
		Whether it is reached by the preferred embedding before
		coexistence requires the separate free-energy selection
		used in Fig.~\ref{fig:equilibriumpeak}.}
	\label{fig:peakgeometry}
\end{figure}
\FloatBarrier
\end{itemize}

\section{Independent coexistence-response checks}
\label{app:analyticchecks}
The following perturbative slope values were recovered from the
archived analytical and numerical verification accompanying an
earlier manuscript draft; they are retained here as a comparison
rather than presented as a newly repeated high-precision run.
The finite-difference estimate uses $[M_t(+\delta)-M_t(-\delta)]/(2b\delta)$
with $\delta=0.002$.
\begin{table}[htbp]
\centering\small
\caption{Einstein-point coexistence displacement
$m_1=\partial(M_t/b)/\partial\lGB|_{T,b,0}$.
The temperature-compensation column equals the difference between
fixed-temperature and fixed-$u_h$ slopes in the stated radial gauge.
The displayed digits are archived numerical estimates, not certified
error bounds.}
\label{tab:restoredslopes}
\begin{tabular}{@{}lrrrrr@{}}\hline
$T/b$&$M_t^{(0)}/b$&Action $m_1$&Direct $m_1$&Fixed $u_h$&Thermal\\\hline
$0.10$&0.492932&0.102878&0.102879&$-0.026917$&0.129795\\
$1/\pi$&0.996177&0.653776&0.653788&$-0.199205$&0.852981\\
$0.50$&1.497160&1.064760&1.064804&$-0.339305$&1.404065\\\hline
\end{tabular}
\end{table}
The largest relative discrepancy in the original multi-step validation
was reported as $0.017\%$. That statement covers the archived numerical
settings and is not a uniform error bound for all parameters.
At these temperatures the fixed-$u_h$ contribution is negative while
the positive thermal term is larger; the decomposition is given analytically in
Eq.~\eqref{eq:slopedecomposition} and discussed in the main text.
\FloatBarrier

\end{document}